\documentclass[prl,twocolumn,a4paper]{revtex4}
\usepackage{graphicx}
\usepackage{bm}
\usepackage[T1]{fontenc}
\usepackage{amsmath,amssymb}
\usepackage{times}
\usepackage{ulem}
\usepackage{color}

\newcommand{\cora}[1]{\textcolor{black}{#1}}

\begin{document}

\title{Spatio-Temporal Imaging of Light Transport in Highly Scattering Media under White Light Illumination}

\author{Amaury Badon}
\author{Dayan Li}
\author{Geoffroy Lerosey}
\author{A. Claude Boccara}
\author{Mathias Fink}
\author{Alexandre Aubry}
 
\affiliation{Institut Langevin, ESPCI ParisTech, PSL Research University, CNRS UMR 7587, 1 rue Jussieu, 75005 Paris, France}


\date{\today}

\begin{abstract}

Imaging the propagation of light in time and space is crucial in optics, notably for the study of complex media. We here demonstrate the passive measurement of time-dependent Green's functions between every point at the surface of a strongly scattering medium by means of low coherence interferometry. The experimental access to this Green's matrix is essential since it contains all the information about the complex trajectories of light within the medium. In particular, the spatio-temporal spreading of the diffusive halo \cora{and the coherent backscattering effect} can be locally investigated in the vicinity of each point acting as a virtual source. On the one hand, this approach allows a quantitative imaging of the diffusion constant in the scattering medium with a spatial resolution of the order of a few transport mean free paths. On the other hand, our approach is able to reveal and quantify the anisotropy of light diffusion, which can be of great interest for optical characterization purposes. This study opens important perspectives both in optical diffuse tomography with potential applications to biomedical imaging and in fundamental physics for the experimental investigation of Anderson localization.

\end{abstract}

\maketitle

\section{Introduction}

Light is the most common probe for investigating complex media at the mesoscopic scale as it offers both an excellent resolution and is non invasive at moderate energies. Nonetheless, due to the inhomogeneous distribution of refractive index, light suffers multiple scattering while propagating in or through the medium. Unveiling the complexity of light scattering \cora{is} then necessary to retrieve the features of an object of interest or of the surrounding environment. In an inhomogeneous medium, it is a classical approach to consider a scattering sample as one realization of a random process, and study statistical physical quantities such as the mean intensity \cite{sheng2006introduction,van1999multiple,akkermans2007mesoscopic}. Under this approach, several physical parameters are relevant to characterize wave propagation in scattering media: the scattering mean-free path $l_s$, the transport mean-free path $l_t$, the diffusion constant $D$, the absorption length $l_a$. Classical back scattering imaging techniques, such as optical coherence tomography, fail when multiple scattering predominates \cite{dunsby2003techniques}. However, one can still measure the long-scale spatial variations of the diffusive parameters. The resulting image is not an image of the refractive index $n(\mathbf{r})$ but e.g., of the diffusion constant $D(\mathbf{r})$ with a resolution of \cora{the order of} the transport mean free path $l_t$, at best. In the literature, diffuse optical tomography is the gold standard technique to reconstruct the spatial distribution of transport parameters at each point of a volume from intensity measurements at the surface \cite{durduran2010diffuse}. Unfortunately, this inverse problem is intrinsically nonlinear with respect to the optical properties of the medium. This method is thus computationally intensive and limited in terms of spatial resolution \cite{puszka,azizi,konovalov} (e.g. 5 mm in human soft tissues).

In this paper, we propose a simple and \cora{efficient} approach that does not require any inversion procedure and provides a direct image of transport parameters with a spatial resolution of the order of a few transport mean free paths. Experimentally, it relies on a passive measurement of time-dependent Green's functions between every point at the surface of a scattering medium without the use of any coherent source. This method is based on the following fundamental result: The cross-correlation (or mutual coherence function) of an incoherent wave field measured at two points A and B can yield the time dependent Green's function $g_{BA}(t)$ between these two points \cite{bart,wapenaar,wapenaar2,snieder2004extracting}. Previously developed in seismology \cite{weller1974seismic,campillo,larose}, in acoustics \cite{weaver,derode} and in the microwave regime \cite{davy}, the Green's function estimation from the cross-correlations of a diffuse wave-field has been recently extended to optics \cite{badon}. \cora{A proof-of-concept experiment has demonstrated the passive measurement of time-dependent Green's functions between individual scatterers using low coherence interferometry. The case of a strongly scattering medium has been also briefly discussed by measuring the autocorrelation signal at one point $A$ of the medium surface. It was shown to converge towards the \textit{self}-Green's function $g_{AA}(t)$ associated with a virtual sensor placed at point $A$.}

\cora{In this paper, this approach is generalized to the passive measurement of time-dependent Green's functions $g_{AB}(t)$ between any point $A$ and $B$ at the surface of a scattering medium. We show in particular how a simple Michelson interferometer allows to simultaneously acquire millions of time-dependent Green's functions under a fully incoherent white light illumination. This non-trivial result means that any point illuminated by a broadband and incoherent light can virtually become a coherent point source as if a femtosecond micro-laser was placed at its location. Reciprocally, it can also become a micro sensor able to measure both the amplitude and phase of light at the femtosecond time scale. Hence, one can measure the impulse response between any of these virtual micro-antennas. This set of responses, that we will refer to as the Green's matrix, contains a wealth of information and provides a unique signature of the complicated trajectories experienced by light in the medium. Its measurement is decisive in many applications, in particular for optical imaging and characterization of complex media.}

\cora{Here, to demonstrate the potential of our approach, we take advantage of the Green's matrix to investigate the spatio-temporal evolution of the intensity in the vicinity of each point acting as a virtual source. This experimental scheme is first applied to the study of a thick layer of $\text{TiO}_2$ nanoparticles, statistically homogenous in terms of disorder. The spatial intensity profile shows two contributions: (i) an incoherent background that results from the incoherent summation of all the multiple scattering paths that the wave can follow within the scattering medium; (ii) a coherent contribution that results from the constructive interference between reciprocal multiple scattering paths, an interference phenomenon known as \textit{coherent backscattering} \cite{kuga1984retroreflectance,PhysRevLett.55.2692,wolf1985weak}. The first component directly accounts for the growth of the diffusive halo within the scattering medium whereas the second component results in a coherent backscattering peak at the source location \cite{larose2004weak,aubry2}. A fit of the incoherent background with a simple diffusion model allows an estimation of the diffusion constant, independently from the absorption losses. The case of a heterogeneous layer made of ZnO nanoparticles is tackled in a second part. The growth of the diffusive halo is investigated over sliding overlapping areas of the scattering medium \cite{aubry2007,aubry2}. It yields an image of the diffusion constant that is shown to be inversely proportional to the concentration of nanoparticles. In a last part, the spatio-temporal evolution of light transport is investigated in a Teflon tape sample. Our approach reveals the anisotropic growth of the diffusive halo induced by the orientation of fibers in the scattering medium.}
\begin{figure*}[htbp]
\includegraphics[width=\textwidth]{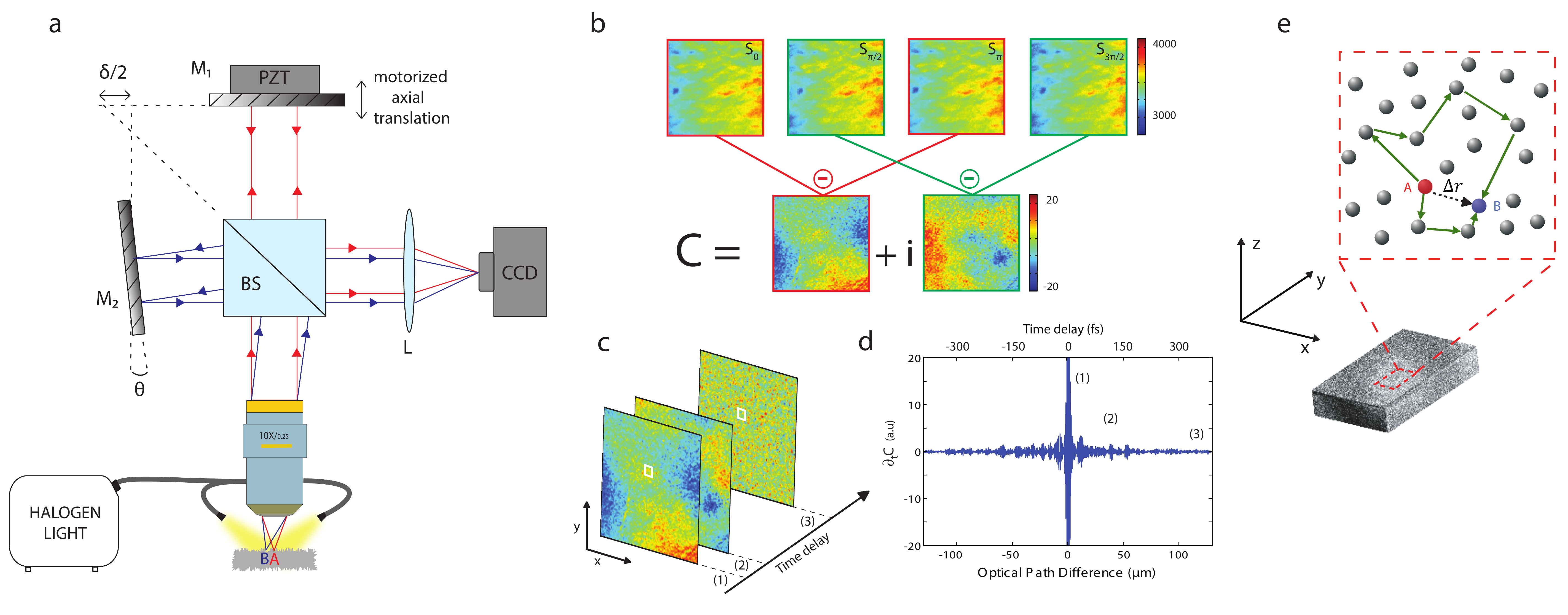}    
\caption{\textbf{Experimental set up and principle of measurement}. (a) Experimental set up: a broadband incoherent light source isotropically illuminates a scattering sample. The spatio-temporal correlation of the scattered wave field is extracted by means of a Michelson interferometer and recorded by a CCD camera. HL: halogen lamp. MO: microscope objective. BS: beam splitter. M: mirror. PZT: piezoelectric actuator. \cora{(b) Application of the ``four phase method'' to the measured intensity patterns $S_{\alpha}(\mathbf{r},\mathbf{r}+\mathbf{\Delta r},t)$ to extract the full-field interferogram $C(\mathbf{r},\mathbf{r}+\mathbf{\Delta r},t)$.} (c) Examples of full-field interferograms acquired at different OPDs \cora{in the TiO$_2$ layer}. \cora{(d) Example of time-derivative of the autocorrelation function $C(\mathbf{r},\mathbf{r},t)$ acquired at a point $\mathbf{r}$ of the CCD camera (marked with the white square in b)}. (e) Principle of our approach: The \cora{time-derivative of the} correlation signal between two points $A$ and $B$ directly provides the difference between the causal and anticausal Green's functions, $g_{AB}(t)-g_{AB}(-t)$, between these two points. This passive measurement mimics an active experiment in which a virtual point-like pulsed source is introduced at point $A$ and a virtual detector records the time-resolved scattered wave-field at point $B$.}
\label{fig1}
\end{figure*}

\section{ Passive measurement of time-dependent Green's functions}

The experimental set up employed for the passive measurement of the point-to-point Green's functions at the surface of a scattering sample is displayed in Fig.~\ref{fig1}. An incoherent broadband light source (650-850 nm) isotropically illuminates a scattering sample. This incident wave-field exhibits a coherence time $\tau_c=10$ fs and a coherence length $l_c = 3$ $\mu$m \cite{badon}.The backscattered wave-field is collected by a microscope objective and sent to a Michelson interferometer, which is here used as a \cora{spatio-temporal} field correlator. The beams coming from the two interference arms are recombined and focused by a lens. A CCD camera conjugated with the sample surface records the output intensity. 
\cora{
\begin{equation}
S_{\alpha}(\mathbf{r},\mathbf{r}+\Delta \mathbf{r},t) =  \int_{0}^T |e^{i\alpha}\psi(\mathbf{r},t+\tau)+\psi(\mathbf{r}+\Delta \mathbf{r},\tau)|^2  \mathrm{d}\tau 
\label{intensity}
\end{equation}
with $\tau$ the absolute time, $\mathbf{r}$ the position vector on the CCD screen, $\psi(\mathbf{r},\tau)$ the scattered wave field associated with the first interference arm, $T$ the integration time of the CCD camera, and $\alpha$ an additional phase term controlled with a piezoelectric actuator placed on mirror $M_1$. The tilt of mirror $M_2$ allows a displacement $\Delta \mathbf{r}$ of the associated wave-field on the CCD camera. The motorized translation of mirror $M_1$ induces a time delay $t = \delta/c$ between the two interferometer arms, with $\delta$ the optical path difference (OPD) and $c$ the light celerity. The interference term is extracted from the four intensity patterns (Eq.\ref{intensity}) recorded at $\alpha=0,$ $\pi/2$, $3\pi/2$ and $\pi$, as depicted in Fig.\ref{fig1}(b)  (``four phase method'' \cite{dubois}). It directly yields the correlation $C$ of the scattered wave-field $\psi$:}
\begin{equation}
\label{cor}
C(\mathbf{r},\mathbf{r}+\Delta \mathbf{r},t)  = \frac{1}{T} \int_{0}^T \psi(\mathbf{r},t+\tau)\psi^*(\mathbf{r}+\Delta \mathbf{r},\tau)  \mathrm{d} \tau \\
\end{equation}
If the incident light is spatially and temporally incoherent, the time derivative of the correlation function $C(\mathbf{r},\mathbf{r+\Delta r},t)$ should converge towards the difference of the causal and anti-causal Green's function \cite{badon}, such that
\begin{equation}
\partial_t C(\mathbf{r},\mathbf{r}+\Delta \mathbf{r},t) \underset{T\rightarrow \infty}{\sim}   g(\mathbf{r},\mathbf{r}+\Delta \mathbf{r},t) -  g(\mathbf{r},\mathbf{r}+\Delta \mathbf{r},-t)
\label{green}
\end{equation} 
where $g(\mathbf{r},\mathbf{r}+\Delta \mathbf{r},t)$ is the time-dependent causal Green's function between a point source at $\mathbf{r}$ and a point detector  $\mathbf{r}+\Delta \mathbf{r}$. This fundamental result means that a passive correlation measurement mimics the following active experiment [see Fig.~\ref{fig1}(d)]: (i) A virtual light source of size $l_c$ located at point $\mathbf{r}$ emits a pulse of duration $\tau_c$; (ii) A virtual point-like receiver located at point $\mathbf{r}+\Delta \mathbf{r}$ records the time-resolved scattered wave-field. This property has been recently demonstrated in optics by retrieving the ballistic and multiple scattering components of the Green's function between individual scatterers \cite{badon}. Following this experimental proof-of-concept, we here investigate the case of strongly scattering media.

\begin{figure*}[htbp]
\includegraphics[width=\textwidth]{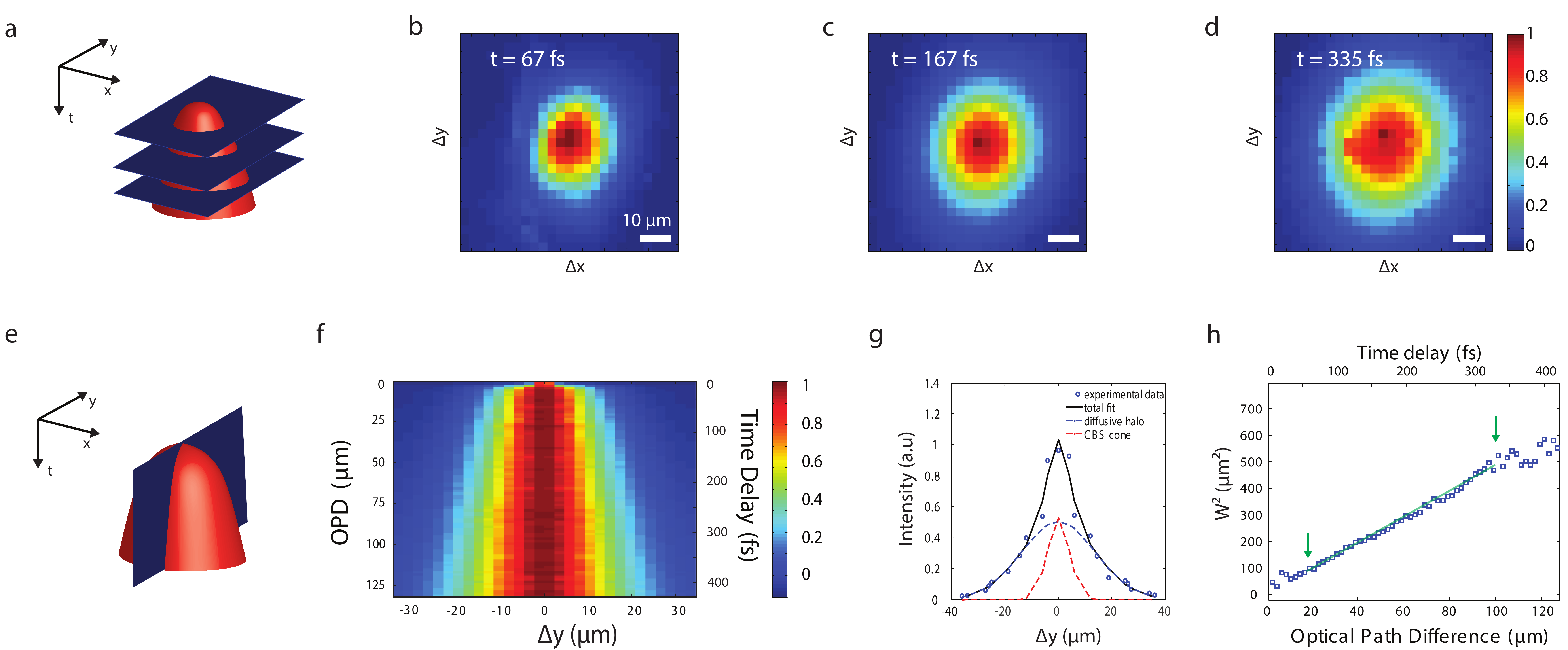}
\caption{ \textbf{Isotropic diffusion from a $\text{TiO}_2$ strongly scattering layer}. (a) Sketch of the spatio-temporal diffuse intensity and its 2D spatial sections for different OPDs or, equivalently, times of flight. (b),(c),(d) Measured spatial distribution of the mean intensity at different OPDs ($\delta = 10$, 50 and 100 µm, respectively). Each intensity map has been renormalized with its maximum. (e) Sketch of the spatio-temporal diffuse intensity and its section for $\Delta x=0$. (f) Measured spatio-temporal diffuse intensity for $\Delta x=0$. The intensity is renormalized by its maximum at each time of flight. (g) Spatial intensity profile (blue disks) obtained at $\delta=$40 $\mu$m fitted with the sum of two gaussian curves (continuous black line) that account for the incoherent intensity (blue dashed line) and the coherent backscattering peak (red dashed line). (h) Time evolution of the mean-square $W^2$ of the diffusive halo (blue squares). A linear fit (green dashed line) \cora{over the OPD range [20 $\mu$m-100 $\mu$m]} (green arrows) yields an estimation of the diffusion constant: $D=730$ $\text{m}^2.\text{s}^{-1}$
}
\label{fig2}
\end{figure*} 
The first sample under study is a 80 $\mu$m-thick layer of $\text{TiO}_2$ nanoparticles.  A set of time-dependent causal and anticausal Green's functions, $ g(\mathbf{r},\mathbf{r}+\Delta \mathbf{r},t) -  g(\mathbf{r},\mathbf{r}+\Delta \mathbf{r},-t)$, has been measured following the experimental procedure described above and depicted in Fig.~\ref{fig1}. \cora{The time derivative of the mutual coherence function, $\partial_t C(\mathbf{r},\mathbf{r}+\Delta \mathbf{r},t)$, is estimated from the finite difference of the mutual coherence function measured at time $t \pm \Delta t/2$, with a time step $\Delta t$=0.66 fs. As the bandwidth of the light source spans from $\lambda_{min}=$400 nm to $\lambda_{max}=$950 nm, the Nyquist criterion is fulfilled ($\Delta t< \lambda_{min}/2c$) and the finite time difference approximation is valid. Fig.~\ref{fig1}(d) displays an example of $\partial_t C(\mathbf{r},\mathbf{r},t)$ in Fig.~\ref{fig1}(d) for an integration time $T=750$ ms. The convergence of $\partial_t C(\mathbf{r},\mathbf{r},t)$  towards the difference between the causal and anticausal Green's function [Eq.\ref{green}] is investigated in Supplement 1. Here, we have access to a satisfying estimation of the Green's function over 100 $\mu$m in terms of OPD, which corresponds to scattering path lengths of approximately 13 $l_t$ [see Sec.3\ref{homo}]. Therefore, the recorded signal contains information on the propagation of light deep into the diffusive regime \cite{peng}. As illustrated by Fig.\ref{fig1}(d), it actually exhibits a long tail due to the numerous scattering events experienced by the wave during its propagation in the medium.}

\section{Spatio-Temporal Imaging of Light Transport}

\subsection{\label{homo}Homogeneously disordered scattering medium}
When one tries to image an unknown medium from a set of Green's functions, an important issue is the importance of multiple scattering relative to single scattering. Multiple scattering is actually a nightmare for classical imaging techniques, which are based on the first Born approximation. However, one can adopt a probabilistic approach and investigate statistical quantities such as the mean intensity\cite{sheng2006introduction}. As the causal and anticausal Green's functions are, by definition, nil for $t<0$ and $t>0$, respectively, the mean intensity can be averaged over the positive and negative times, such that:
\begin{eqnarray}
& &  I(\mathbf{\Delta r}, t )  \nonumber \\
& & =  \left \{ \left \langle \left | \partial_t C(\mathbf{r},\mathbf{r}+\Delta \mathbf{r},t) \right |^2 \right \rangle +\left \langle \left |\partial_t C(\mathbf{r},\mathbf{r}+\Delta \mathbf{r},-t) \right |^2 \right \rangle \right \}  \nonumber \\
\label{eq1}
& &  = \left \{ \left \langle \left | g(\mathbf{r},\mathbf{r+\Delta r}, t ) \right |^2 \right \rangle + \left \langle \left | g(\mathbf{r},\mathbf{r+\Delta r}, -t ) \right |^2 \right \rangle \right \} 
\end{eqnarray}
where the symbol $\langle . \rangle$ denotes an average over the CCD pixels, i.e over the position vector $\mathbf{r}$ of the virtual source on the sample surface. As we shall see, the evolution of this quantity as a function of the source-receiver relative position $\mathbf{\Delta r}$ and the time of flight $t$ allows to describe quantitatively the transport of light through the scattering medium under study.

Fig.~\ref{fig2} displays the spatio-temporal evolution of the mean intensity deduced from the set of point-to-point Green's functions measured in the $\text{TiO}_2$ layer. \cora{Note that the noise background is priorly subtracted from the measured intensity (see Supplement 1).} Each spatial intensity profile is then renormalized by its maximum at each time of flight $t$. Figure~\ref{fig2}(b,c,d) displays the result at different OPDs ($\delta=$10, 50 and 100 µm, respectively). Not surprisingly, we retrieve the feature of a diffusive halo whose spatial extent increases with time at the same speed in all directions. This isotropic growth is easily accounted for by the fact that the $\text{TiO}_2$ nanoparticles are small compared to the wavelength, randomly packed and display an isotropic differential cross-section.

Now that we have observed qualitatively the diffusive spreading of the mean intensity, we aim to precisely characterize the scattering properties of the medium with a quantitative measurement of the diffusion constant $D$. Theoretically, the multiple scattering intensity can be split into two terms. First, the incoherent contribution, denoted as $I_{inc}(\Delta r,t)$, corresponds to the interference of the wave with itself. In the deep multiple scattering regime, this term obeys the \cora{diffusion equation  [See \href{link}{Supplement 1}]. In the time domain, the corresponding reflection profile has been derived by Patterson \textit{et al.} \cite{Patterson}}:
\begin{equation} 
\label{eq1b}
I_{inc}({\Delta r },t)= I_z(t) . \exp\left(-\frac{{{\Delta r} }^2}{4Dt}\right)
\end{equation}
with $D$ the diffusion constant and $I_z(t)$ a physical quantity that depends on the sample thickness $L$, the transport mean-free path $l_t$, the absorption length $l_a$ but not on the transverse coordinate $\mathbf{r}$ [See \href{link}{Supplement 1}]. Secondly, the coherent contribution, denoted $I_{coh}(\Delta r,t)$, corresponds to the interference of the wave with its reciprocal counterpart \cite{kuga1984retroreflectance,PhysRevLett.55.2692,wolf1985weak} and is expressed as \cite{aubry2007}: 
\begin{equation} 
\label{eq12}
I_{coh}({\Delta r },t)= I_z(t) . \exp\left(-\frac{{{\Delta r} }^2}{l_c^2}\right)
\end{equation}
At a given time of flight, the normalized multiple scattering intensity is thus given by:
\begin{equation} 
\label{eq7}
\frac{I_{MS}({\Delta r },{t)}}{I_{MS}(0,{t)}}= \frac{1}{2}{\left[\exp\left(-\frac{{{\Delta r} }^2}{4Dt}\right)+\exp\left(-\frac{{{\Delta r} }^2}{l^2_c}\right)\right]}
\end{equation}
In our configuration,  the spatial distribution of the intensity at a given time has the following shape [see Fig.~\ref{fig2}(g)]: a narrow, steep peak (the  coherent  contribution), on top of a wider pedestal that widens with time (the incoherent contribution). It is worth noting that due to the normalization process, the multiple scattering intensity profile does not depend on the absorption or the thickness of the sample. By fitting this intensity profile at each time of flight with two gaussian curves, one can separate the coherent and incoherent contributions [see Fig.~\ref{fig2}(g)]. The width $W$ of the incoherent background directly accounts for the spatial extent of the diffusive halo. Figure~\ref{fig2}(d) displays the evolution of $W^2$ versus time. From Eq.\ref{eq7}, we see that the slope of $W^2$ versus time should be equal to $2D$. The linear fit of $W^2$ in Fig.~\ref{fig2}(d) allows to measure the diffusion constant in the $\text{TiO}_2$ scattering sample: we find $D=730$ $\text{m}^2.\text{s}^{-1}$.  \cora{Note that the linear fit is applied from an OPD of 20 $\mu$m, which corresponds to a typical penetration depth of more than  3 $l_t$. Before this time, the diffusion approximation is not valid yet and the evolution of $W^2$ is not linear. }

\cora{In Supplement 1, the measured value of $D$ is compared to the result of a more conventional time-of-flight experiment which, unlike our approach, is sensitive to absorption losses. The combination of both methods allows to estimate quantitatively the absorption length: $l_a \simeq 118$ $\mu$m. Our measurement is in qualitative agreement with transmission time-of-flight measurements performed in other TiO$_2$ samples \cite{maret_paint}.} An estimation of the transport mean-free path $l_t$ is also possible using the relation $D=v_e l_t/3$, with $v_e$ the energy velocity. The size of the nanoparticles being small compared to the wavelength and the medium being diluted, we can consider the energy velocity $v_e$ to be close to the vacuum light velocity $c$ \cite{van1991speed}. Thus, we obtain the following qualitative estimation for the transport mean free path: $l_t \sim 7.5$ $\mu$m.

The access to transport parameters from the correlation of a diffuse wave-field is a precious information for the optical characterization of a scattering layer. However, this measurement is averaged over the whole field-of-view (FOV) and does not provide any local information on disorder. This issue is tackled in the next subsection.
\begin{figure*}[htbp]
\includegraphics[width=\textwidth]{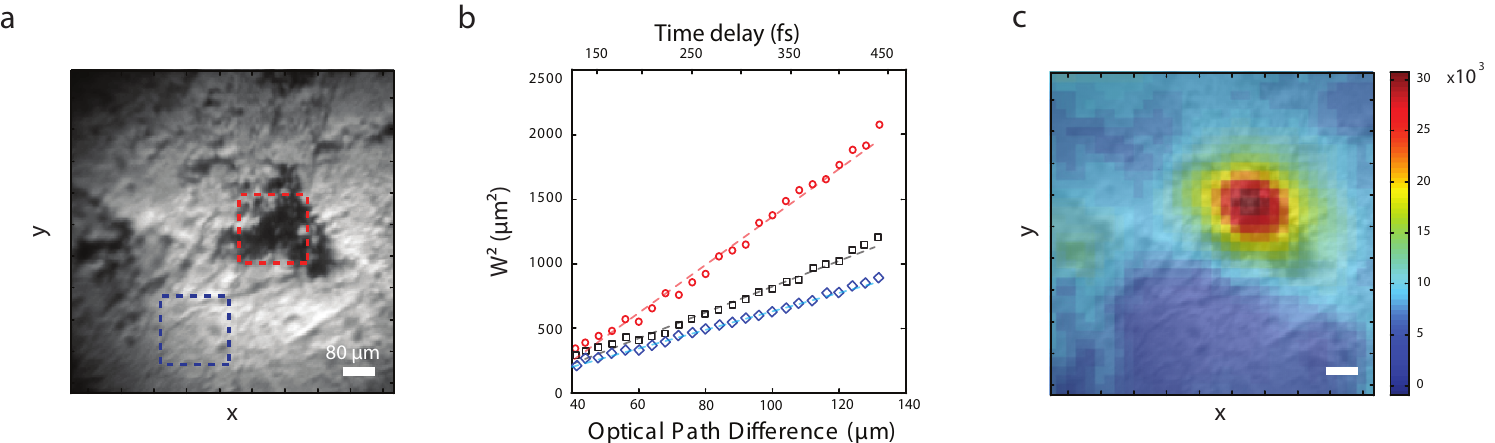}
\caption{\textbf{Imaging of the diffusion constant in a scattering layer heterogeneous in disorder}.  (a) Microscopic image of the \cora{ZnO} scattering layer. (b)Time evolution of the mean-squared width of the diffusive halo averaged over the whole sample surface (black squares), over areas surrounded by the blue dashed square (blue diamonds) and the red dashed square (red circles) in (a). A linear fit of each experimental data set (dashed lines) \cora{over the OPD range [40 $\mu$m-130 $\mu$m]} yields an estimation of the diffusion constant in the corresponding areas. (c) Superimposition of the measured diffusion constant map onto the sample image. The color scale is in $\text{m}^2.\text{s}^{-1}.$}
\label{fig4}
\end{figure*}

\subsection{Heterogeneously disordered scattering medium}

Until now, the disorder was statistically homogeneous in the scattering sample under study. We now investigate the case of a medium with a space-dependent concentration of scatterers. For this purpose we have synthesized a heterogeneously disordered layer of ZnO nanoparticles. Variations in the concentration of scatterers are visible at a scale smaller than 100 $\mu$m in the microscopic image [see Fig.~\ref{fig4}(a)]. Cross-correlations of the diffuse wave-field are recorded using the procedure described in Sec.~2. \cora{A satisfying estimation of the Green's function is found over 130 $\mu$m in terms of OPD and the noise background is subtracted from the measured mean intensity prior to data analysis}. The spatial extent $W$ of the diffusive halo is estimated at each time of flight by fitting the normalized multiple scattering intensity profile [Eq.~\ref{eq7}] with two gaussian curves. First, the mean intensity is averaged over all the active pixels of the CCD. A linear fit of $W^2$ versus time [Fig.~\ref{fig4}(b)] allows an estimation of the diffusion constant average over a FOV of 5 $\text{mm}^2$: $<D(\mathbf{r})>=1425$ $\text{m}^2.\text{s}^{-1}$. The FOV being at least one order of magnitude larger than the typical scale at which the concentration of scatterers fluctuates, such a measurement does not provide a satisfying characterization of the disorder in the scattering sample.  

To cope with this issue, a local approach is needed. Instead of considering the mean intensity over the whole sample surface, one can average the intensity over sub-areas of the CCD camera. For instance, one can consider sub-areas delineated with red and blue squares in Fig.~\ref{fig4}(a) that display low and high concentration of scatterers, respectively. The corresponding diffusive halos show a contrasted speed of expansion [see Fig.~\ref{fig4}(b)]. Quantitatively, a different diffusion constant is measured in both areas. We find $D=1125 \text{m}^2.\text{s}^{-1}$ in the blue area and $D=2770 \text{m}^2.\text{s}^{-1}$ in the red area. This is in agreement with the fact that a higher concentration of scatterers implies a smaller diffusion constant. $D$ is actually proportional to the transport mean free path $l_t$ which itself scales as the inverse of the concentration of scatterers \cite{van1991speed}. Hence, the diffusion contrast highlighted by Fig.~\ref{fig4}(b) indicates that the blue area is twice and half more concentrated than the red one.

A further development is to build an image of the medium from local measurements of the diffusion constant. This is done by considering the mean intensity over sliding windows of $100\times100$ $\mu \text{m}^2$. The spatial extent $W^2$ of the diffusive halo is fitted linearly over time and yields an estimation of the diffusion constant at the center of each window. The map of the diffusion constant is superimposed to the reflectivity image of the scattering medium in Fig.~\ref{fig4}(c). A qualitative agreement is found between the two images since the diffusion constant is larger in areas where the concentration in scatterers is lower. However, both images bring different information since the microscope image is only related to the concentration of scatterers at the surface of the medium whereas the diffusion constant also depends on the nature of disorder below the surface. A 3D image could thus be built from a passive Green's functions retrieval but it would require inversion schemes like in optical diffuse tomography \cite{durduran2010diffuse}.

\cora{The spatial resolution $\delta r$ of the image displayed in Fig.~\ref{fig4}(c) is basically limited by the transverse spreading of the diffusive halo, such that
\begin{equation}
\delta r \sim \sqrt{Dt_{max}}
\end{equation}  
where $t_{max}$ is the largest time delay investigated in the data analysis. Here $t_{max}=450$ fs, hence $\delta r$ is of the order of 25 $\mu$m. Considering a range of shorter time delays would improve the resolution but, in the mean time, would limit the precision of our measurement. Note that the resolution of the image displayed in Fig.~\ref{fig4}(c) is also limited by the size of the spatial sliding window. A better resolution would be obtained if a smaller sliding window was considered. But, in this case, the average of the intensity would not be satisfying. Residual fluctuations of the intensity pattern would be too high because of the lack of average over disorder configurations. Consequently, a compromise has to be found between the size of the sliding window (\textit{i.e} the image resolution) and a sufficient average over disorder configurations (\textit{i.e} the signal-to-noise ratio).}

\subsection{Anisotropic scattering medium}

\begin{figure*}[htbp]
\includegraphics[width=\textwidth]{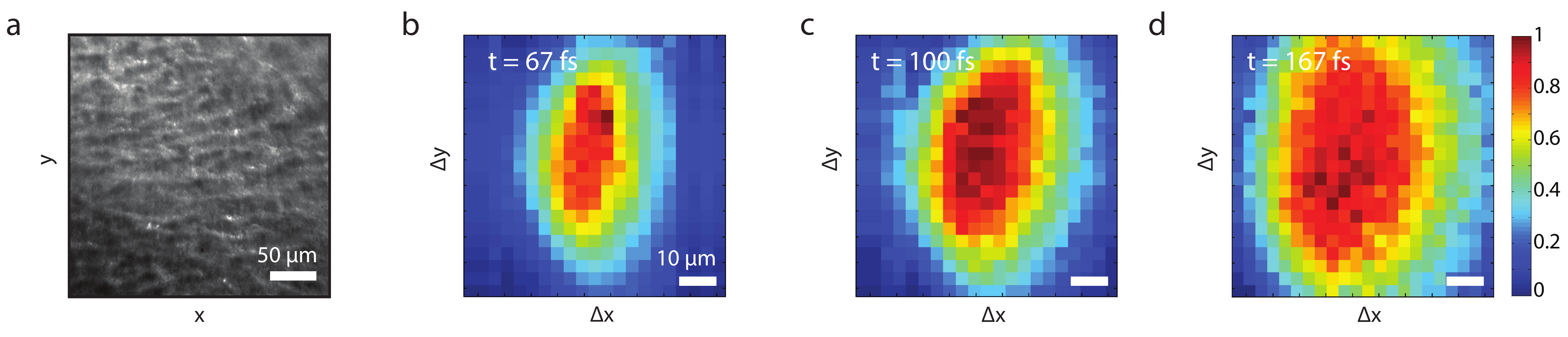}
\caption{\textbf{Anisotropic diffusion in a stretched Teflon tape}.(a) Microscopic image of the sample. (b),(c),(d) Spatial intensity distribution recorded at different OPDs ($\delta=$20, 30 and 50 µm, respectively). Each intensity has been renormalized by its maximum.}
\label{fig3}
\end{figure*}
Until now, we have only considered the case of isotropic scattering media. However, many complex samples such as biological tissues \cite{nickell}, nematic crystals \cite{wiersma_nematic,van_tiggelen_nematic}, porous materials \cite{johnson2} or fibrous media \cite{simon} give rise to anisotropic light diffusion. In such media, the scalar diffusion constant should be replaced by a diffusion tensor [See \href{link}{Supplement 1}]. Providing that the principal directions of anisotropy are aligned with the Cartesian coordinates, this tensor is diagonal \cite{konecky2011imaging}. The three non zero coefficient $D_{xx}$, $D_{yy}$ and $D_{zz}$ correspond to the diffusion constant along the directions $x$, $y$ and $z$, respectively. 
To illustrate the ability of our approach in measuring different components of the diffusion tensor, we now investigate light diffusion in a piece of thread seal tape (Polytetrafluoroethylene) which is made of fibers aligned along the x-direction [see Fig.~\ref{fig3}(a)]. Strong forward scattering arises along the normal to the fibers' direction. Hence, light diffusion is supposed to be slower along the fibers: $D_{xx} < D_{yy}$. To check that hypothesis, the experimental procedure described in Sec.~2 is performed. The measured Green's matrix allows to investigate the spatio-temporal evolution of the mean intensity [Eq.~\ref{eq1}] averaged over the whole FOV. Fig.~\ref{fig3} displays the diffusive halo in the $x-y$ plane at different times of flight. The intensity distribution clearly presents an elongated shape along the $y-$direction. As expected, light diffusion is thus slower along the $x-$direction. This confirms the anisotropy of the sample and the ability of our approach to reveal it.

In Supplement 1, the growth of the diffusive halo is quantitatively investigated both in the $x-$ and $y-$directions. The components of the diffusion tensor are estimated following the procedure described in Sec.~3A.  We find $D_{xx}=1100$ $\text{m}^2.\text{s}^{-1}$ and $D_{yy}=3000$ $\text{m}^2.\text{s}^{-1}$. An anisotropy factor $g$ can be deduced from the measurement of the diffusion tensor in the $x-y$ plane \cite{alerstam}, such that $g=1-\left(D_{xx}/D_{yy}\right)=0.63$. The anisotropy of diffusion is thus successfully quantified by our approach.

\section{Discussion}
The first point we would like to emphasize is the potential impact of our approach in optical diffuse tomography. \cora{In contrast with most studies in the literature that only involve intensity measurements between a limited number of active sources and detectors, our method allows to passively measure a whole set of time-dependent responses between every point at the surface of a scattering medium. Moreover, compared to similar recent experiments performed in transmission with a coherent illumination scheme \cite{pattelli,sperling2013direct}, backscattering measurements open the possibility to address the very first scattering events that may reveal fine structural features hardly accessible otherwise. The access to short times of flight enables a local study of the growth of the diffusive halo and a direct two-dimensional imaging of transport parameters with a resolution not given by the thickness of the sample, as it would be in transmission, but of the order of the transport mean free path. At last, our approach is by no means limited to 2D imaging. One perspective of this work is actually to apply inversion schemes of the diffusion or radiative transfer equations \cite{arridge2009optical} to recover a 3D image of the scattering medium. }

A passive measurement of the Green's matrix can also enable a great leap forward in fundamental physics. In this work, we have shown, for instance, how the anisotropy of light diffusion can be revealed and quantified. A further step would be to investigate exotic transport phenomena such as super- and sub-diffusion of light \cite{bertolotti2010engineering,barthelemy2008levy} or, more ambitiously, to provide a direct proof for Anderson localization \cite{anderson1958absence,lagendijk2009fifty}. Past experimental studies of Anderson localization in optics have been ambiguous, due either to absorption effects \cite{wiersma1997localization,storzer2006observation} or fluorescence issues \cite{maret_paint}. On the contrary, our approach would remove any possible ambiguity. Space- and time-dependent experiments can enable absorption-independent observations of localization \cite{hu2008localization}. Moreover, a white light illumination in a backscattering configuration limits the incident power required for a satisfactory signal-to-noise ratio, contrary to a transmission configuration that implies a much less energetic signal through a thick scattering layer ($L>>l_t$). Our experimental scheme would thus prevent from fluorescence or non-linear phenomena \cite{maret_paint}. It also allows to investigate phenomena like coherent backscattering \cite{cherroret,cobus2016} or recurrent scattering \cite{aubry_2014} that constitute relevant observables at the onset of Anderson localization

\section{Conclusion}

In summary, we have proposed a novel approach to investigate light transport through complex media. Inspired by previous studies in ultrasound and seismology, we have shown how a set of time-dependent Green's functions can be passively extracted between every point at the surface of a scattering medium placed under white light illumination. The experimental access to this Green's matrix is essential since it contains all the information about the complex trajectories of light within the medium. Adopting a probabilistic and local approach, one can, for instance, investigate the spatio-temporal expansion of the diffusive halo in different parts of the sample. On the one hand, this allows a two-dimensional imaging of the diffusion constant at the surface of the scattering medium, with a spatial resolution of the order of few transport mean free paths. On the other hand, our approach is able to reveal and quantify the anisotropy of light diffusion, which can be of great interest for optical characterization purposes.  This work opens important perspectives both in optical diffuse tomography with potential applications to biomedical imaging and in fundamental physics with the quest for an experimental demonstration of Anderson localization in 3D.

\section*{Funding Information} 
LABEX WIFI (Laboratory of Excellence within the French Program Investments for the Future, ANR-10-LABX-24 and ANR-10-IDEX-0001-02 PSL*); European Research Council (ERC Synergy HELMHOLTZ); ``Direction Générale de l'Armement''(DGA);  Agence Nationale de la Recherche (ANR-14-CE26-0032 LOVE); High Council for Scientific and Technological Cooperation between France and Israel (P2R Israel N$^o$ 29704SC)

\section*{Acknowledgments} 
We thank O. Katz for fruitful discussions.

\newpage

\renewcommand{\thefigure}{S\arabic{figure}}
\renewcommand{\theequation}{S\arabic{equation}}
\renewcommand{\thetable}{S\arabic{table}}
\setcounter{figure}{1} 

\begin{center}
\Large{\bf{Supplementary Material}}
\end{center}
\normalsize

This document provides supplementary information on the convergence of the time derivative of the mutual coherence function towards the Green's functions, the measurement of the diffusion constant components in isotropic and anisotropic scattering media and its comparison with a time-of-flight experiment.

\section{Convergence of the time derivative of the mutual coherence function}
\begin{figure}[ht!]
\begin{center}
\includegraphics[width=8cm]{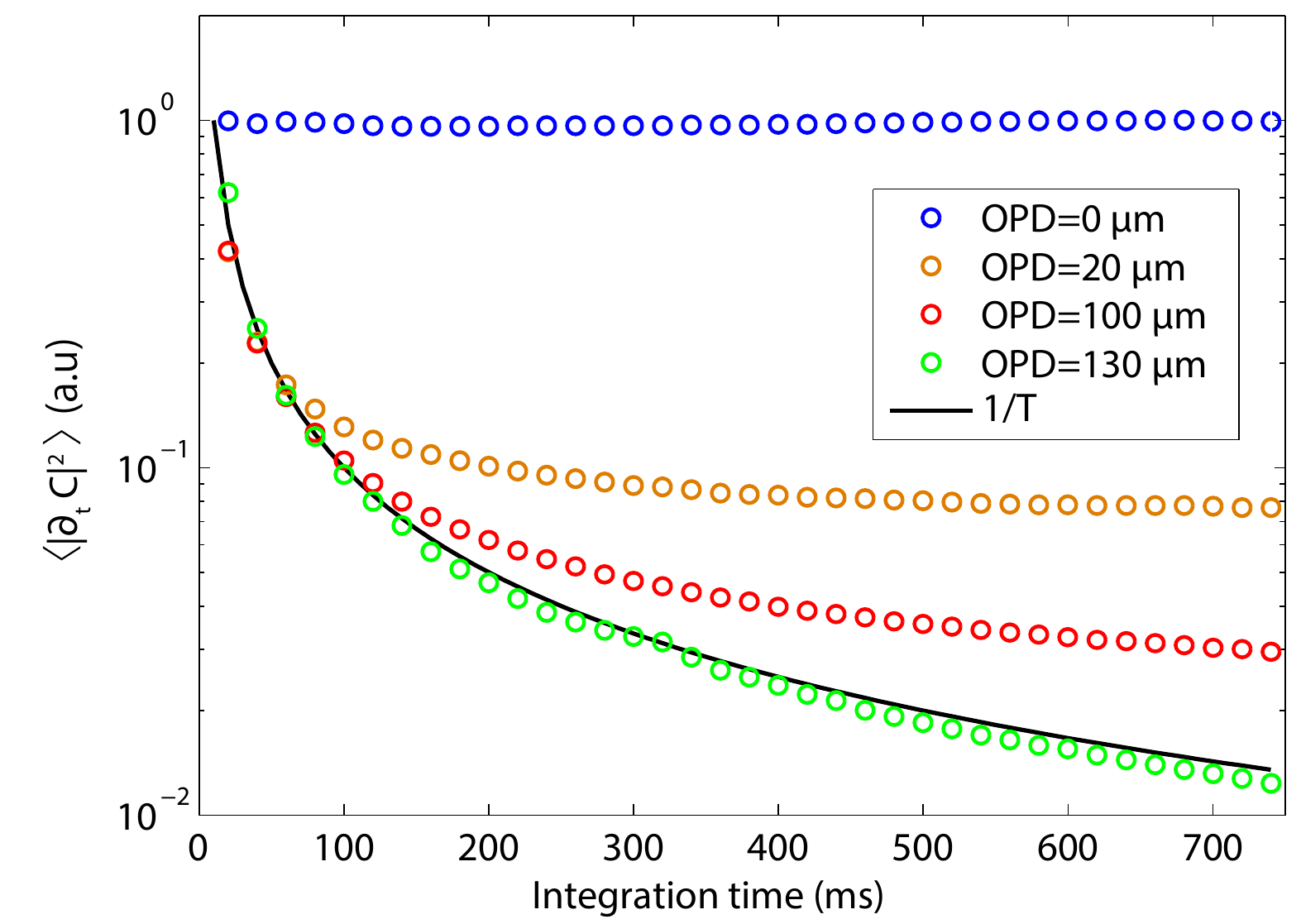}    
\caption{\textbf{Convergence of the time derivative of the mutual coherence function in the TiO$_2$ layer}. The spatially averaged intensity of the mutiual coherence functoin, $\left \langle |\partial_t C|^2 \right \rangle $, is plotted as a function of the integration time $T$ at different OPDs. The continous black line shows for comparison the expected noise intensity decrease as $1/T$ (see Eq.\ref{conv3}).}
\label{convergence}
\end{center}
\end{figure}

The passive measurement of the Green's function is based on the fundamental property that the time derivative of the correlation function should converge for infinite integration times towards the difference between the causal and anti-causal Green's functions [see Eq.1 of the accompanying paper]. In practice, this integration time remains fixed to $T=750$ ms in the experiments shown in the accompanying paper. Hence the measured signal may be polluted by noise. To assess the signal-to-noise ratio (SNR) in our measurements, the convergence of $\partial_t C$ as a function of the integration time $T$ should be investigated. It can be expressed as the sum of a deterministic term that resists to average, the expected Green's functions, and a noise term that should vanish with average
\begin{equation}
\label{conv1}
\partial_t C (t) = \underbrace{\left \lbrace g(t) -  g(-t) \right \rbrace}_{\mbox{signal}} + \underbrace{\frac{1}{T} \int_0^T n(\tau) d\tau}_{\mbox{noise}}
\end{equation}
where $n(\tau)$ accounts for the incoherent noise term whose coherence time $\tau_c$ is governed by the bandwidth of the white light source. To estimate the SNR, the mean intensity of Eq.\ref{conv1} should be considered. It yields
\begin{equation}
\label{conv2}
\left \langle |\partial_t C|^2 \right \rangle =  \underbrace{  \left \langle \left | \left \lbrace g(t) -  g(-t) \right \rbrace \right |^2\right \rangle}_{\mbox{signal intensity}} + \underbrace{\frac{1}{T^2} \left | \left \langle \int_0^T n(\tau) d\tau \right \rangle  \right |^2}_{\mbox{noise intensity}}
\end{equation}
The coherence properties of the noise $n(\tau)$ allows to simplify the last equation into
\begin{equation}
\label{conv3}
\left \langle |\partial_t C|^2 \right \rangle =  \underbrace{  \left \langle \left | \left \lbrace g(t) -  g(-t) \right \rbrace \right |^2\right \rangle}_{\mbox{signal intensity}} + \underbrace{\frac{\tau_c}{T} \left \langle  \left |  n(\tau) \right |^2  \right \rangle}_{\mbox{noise intensity}}
\end{equation}
Not surprisingly, the noise intensity should decrease as the inverse of the integration time $T$. Fig.\ref{convergence} confirms this behavior by showing the spatially averaged intensity of $\partial_t C (t) $ measured in the TiO$_2$ layer as a function of $T$. 
For short time delays ($\delta = 0 $ $\mu$m in Fig.\ref{convergence}), the signal intensity is sufficiently large to make the convergence of $\partial_t C(t)$ nearly immediate. For intermediate time delays ($\delta = 20-100 $ $\mu$m in Fig.\ref{convergence}), $\left \langle |\partial_t C|^2 \right \rangle $ decreases as $1/T$ for small integration times before saturating at the signal level for large integration times. This means that the integration time $T=750$ ms is enough to get rid of most of the noise background and have a satisfying estimation of the Green's function. On the contrary, for larger time delays ($\delta=130$ $\mu$m in Fig.\ref{convergence}), the signal intensity is too weak to emerge from the noise background: $\left \langle |\partial_t C|^2 \right \rangle $ decreases as $1/T$ over the whole integration time range, which is characteristic of a predominant noise background. The value of $\left \langle |\partial_t C|^2 \right \rangle $ at time $T=750$ ms allows to determine the noise level in our experiment. A SNR can be derived for each OPD: SNR$\sim 10^2$ at $\delta=0$ $\mu$m, SNR$\sim 8$ at $\delta=20$ $\mu$m and SNR$\sim 3$ at $\delta=100$ $\mu$m. Note that this noise background is systematically subtracted from the raw mean intensity prior to investigating the growth of the diffusive halo in each experiment.

\section{Incoherent intensity in the multiple scattering regime}

\begin{figure}[ht!]
\begin{center}
\includegraphics[width=5cm]{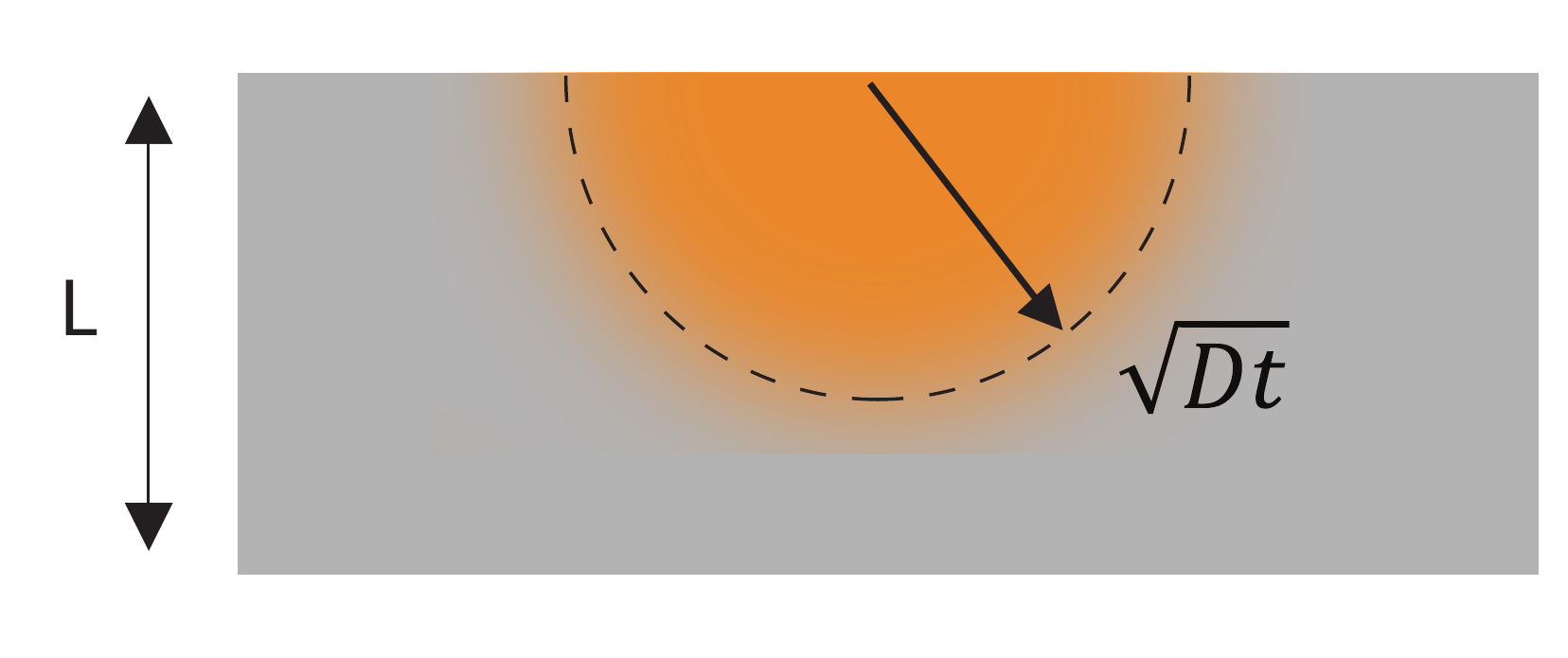}    
\caption{\textbf{Expansion of the diffusive halo in a scattering slab}}
\label{S0}
\end{center}
\end{figure}

In this section we investigate the case of the propagation of light in the multiple scattering regime. As mentioned in the article, the energy density $W$ whose flux corresponds to the incoherent intensity obeys the diffusion equation :
\begin{equation}
\frac{\partial W(\mathbf{r},t)}{\partial t} = D \Delta  W(\mathbf{r},t)-\frac{c}{l_a}W(\mathbf{r},t) + S(\mathbf{r},t).
\label{isotrope}
\end{equation}
with $S(\mathbf{r},t)$ a source function, $l_a$ the absorption mean free path and $D$ the diffusion constant. The solution of this equation corresponds to a diffusive halo whose spatial extent broadens with time [see Fig.~\ref{S0}]. Its mathematical expression also depends on the boundary conditions. In the accompanying paper, the samples under study consist in strongly scattering layers of thickness $L\sim$ 80 $\mu$m. The characteristic time for the diffusive wave-field to travel through the scattering medium is given by the Thouless time $\tau_D=L^2/D \sim 6$ ps for $D=730$ $\text{m}^2/\text{s}$. As the temporal range investigated in our study is of the order of 500 fs, the scattering medium can thus be considered as semi-infinite.
In these conditions, \textit{Patterson et al.} \cite{Patterson} established the expression for the backscattered incoherent intensity :
\begin{equation}
I_{inc}(\Delta r,t)=(4\pi D)^{-3/2}z_0 t^{-5/2} \text{exp}\left(-\frac{ct}{l_a} \right) \text{exp}\left(-\frac{\Delta r^2}{4Dt}\right)
\label{patterson}
\end{equation}
with $z_0$ the extrapolation length \cite{Durian} and $\Delta r$ the source-receiver distance. The incoherent intensity can thus be written as the product of two terms:
\begin{equation} 
\label{}
I_{inc}(\Delta r,t)= I_z(t) \exp\left(-\frac{\Delta r^2}{4Dt}\right)
\end{equation}
with :
\begin{equation} 
\label{}
I_z(t)= (4\pi D)^{-3/2}z_0 t^{-5/2} \text{exp}\left(-\frac{ct}{l_a} \right)
\end{equation}
The first term $I_z(t)$ accounts for the temporal decreasing of the incoherent intensity whereas the second term in $\exp(-\Delta r^2/4Dt)$ accounts for the temporal growth of the diffusive halo. By normalizing the incoherent intensity at each time flight, one can therefore investigate the the diffusive properties of the scattering medium independently from the absorption losses.

\section{Incoherent intensity for an anisotropic scattering medium}

In this section, we investigate the case of the propagation of light in an anisotropic scattering medium. In such a medium, the diffusion equation is now given by \cite{wiersma_nematic}:
\begin{equation}
\frac{\partial W(\mathbf{r},t)}{\partial t} = \mathbf{\nabla. D \nabla} W(\mathbf{r},t)-\frac{c}{l_a}W(\mathbf{r},t) + S(\mathbf{r},t).
\label{anisotrope}
\end{equation}
with $W(\mathbf{r},t)$ the energy density, $S(\mathbf{r},t)$ a source function, $l_a$ the absorption mean free path and $\mathbf{D}$ the diffusion tensor. According to Ref.\cite{simon}, the corresponding incoherent intensity is given by:
\begin{equation}
I(\Delta x,\Delta y,t)  = I_z(t) \text{exp}\left(-\frac{\Delta x^2}{D_{xx}}\right)\text{exp}\left(-\frac{\Delta y^2}{D_{yy}}\right). 
\label{anisotrope2}
\end{equation}
with $D_{xx}$ and $D_{yy}$ the in-plane components of the diffusion tensor. $\Delta x$ and $\Delta y$ are the projections of the source-receiver relative positions along the $x$ and $y$-axis. For a semi-infinite medium, $I_z$ is given by:
\begin{equation}
I_z(t)= \left(D_{xx}D_{yy}D_{zz}\right)^{-1/2} z_0 t^{5/2} \text{exp}\left(-\frac{ct}{l_a} \right).
\end{equation}
At a given time of flight, the normalized incoherent intensity can be expressed as:
\begin{equation}
\frac{I_{inc}(\Delta x,\Delta y,t)}{I_{inc}(0,0,t)}  =  \text{exp}\left(-\frac{\Delta x^2}{4D_{xx}t}\right)\text{exp}\left(-\frac{\Delta y^2}{4D_{yy}t}\right)
\end{equation}
As in the isotropic case, the normalized incoherent intensity is independent on the absorption losses. 

\begin{figure}[htbp]
\includegraphics[width=8cm]{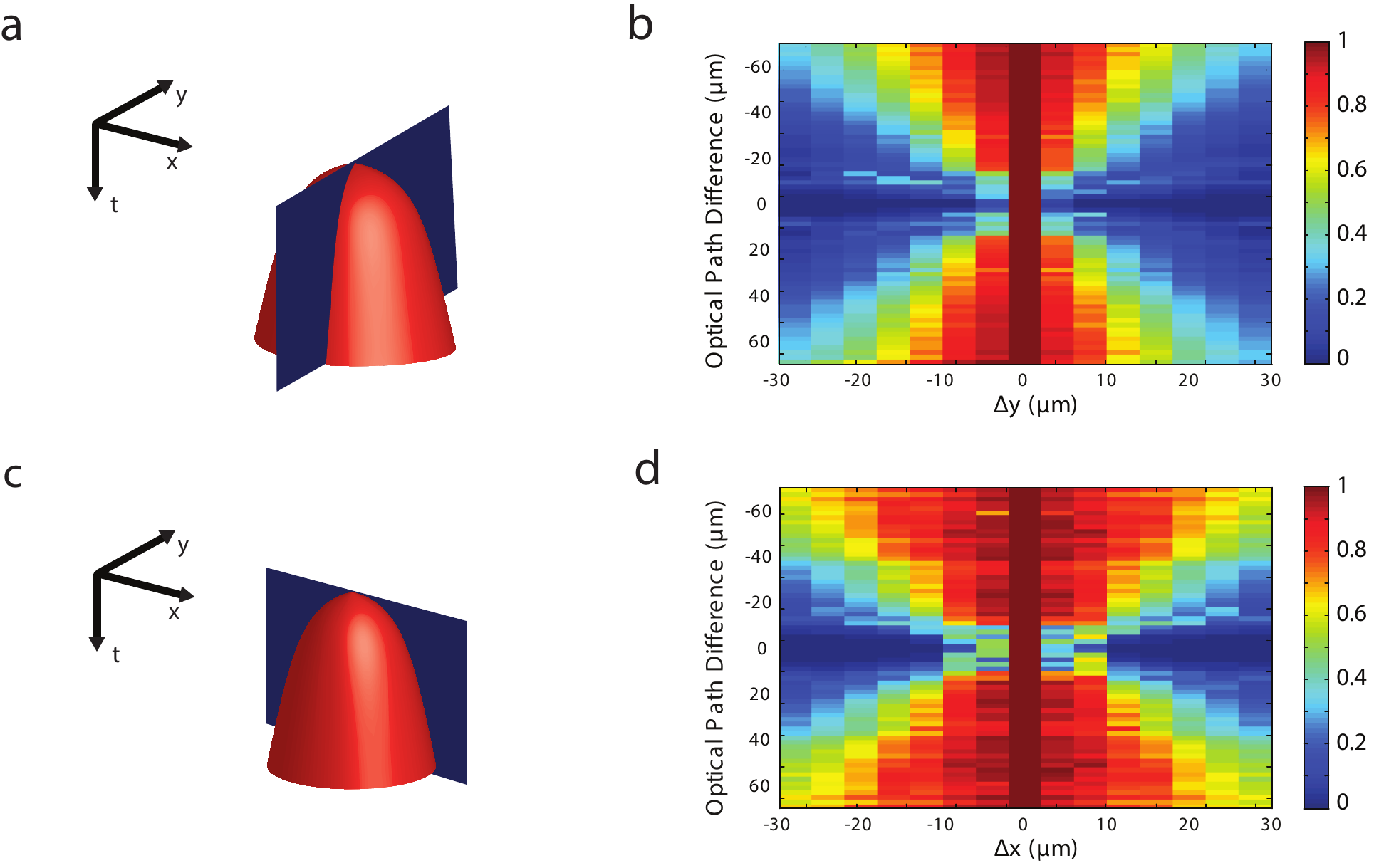}    
\caption{\textbf{Anisotropic diffusion in a stretched Teflon tape}. (a) Sketch of the spatio-temporal diffusive halo and its section at $\Delta x=0$. (b) Measured spatio-temporal evolution of the mean intensity at $\Delta x=0$. The intensity is renormalized by its maximum at each time of flight. (c) Sketch of the spatio-temporal evolution of the mean intensity and its section at $\Delta y=0$. (d) Measured spatio-temporal evolution of the mean intensity for $\Delta y=0$. The intensity is renormalized by its maximum at each time of flight.}
\label{S1}
\end{figure}

Fig.~\ref{S1} displays the spatio-temporal evolution of the mean intensity measured in the stretched Teflon tape. The two components $D_{xx}$ and $D_{yy}$ of the diffusion tensor can be estimated separately. First, the relative position  $\Delta x$ is set to 0 and the mean intensity is investigated as a function of time and $\Delta y$ [see Fig.~\ref{S1}(b)]. A linear fit of the square width $W^2$ of the diffusive halo versus time allows an estimation of $D_{yy}$=3000 $\text{m}^2.\text{s}^{-1}$. Alternatively, by setting $\Delta y=0$, we can investigate the diffuse intensity as a function of time and $\Delta x$  [see Fig.\ref{S1}(d)]. Again, a linear fit of $W^2$ yields an estimation of $D_{xx}$=1100 $\text{m}^2.\text{s}^{-1}$.

\section{Time-of-flight distribution in the TiO$_2$ layer}
In this section, we investigate the time-of-flight distribution of the reflected intensity integrated over the surface of the scattering sample, 
\begin{equation}
R(t)=\int I(\Delta \mathbf{r},t) d^2 \Delta \mathbf{r}.
\end{equation}
The decay of the reflected intensity with time bears particular signatures of light diffusion. For times of flight smaller than the Thouless time $\tau_D $, the medium can be considered as semi-infinite. In that case, a power law decay is expected for the reflected intensity at the surface of the sample \cite{Johnson}:
\begin{equation}
\label{power}
R(t) \propto t^{-3/2}. 
\end{equation}
At longer times of flight, the finite sample size should come into play so that the decay becomes similar to that of time-dependent transmission, i.e., exponential in the diffuse regime \cite{Johnson,page2}, such that
\begin{equation}
\label{exp}
R(t)\propto \exp \left ( - \frac{ct}{l_a} \right ) \exp \left ( - \frac{\pi^2 D t}{(L+2z_o)^2} \right ) 
\end{equation}
In absence of absorption ($l_a \rightarrow \infty $), the time-of-flight distribution of the backscattered intensity is thus an alternative way of measuring the diffusion constant $D$. 

\begin{figure}[ht!]
\begin{center}
\includegraphics[width=8cm]{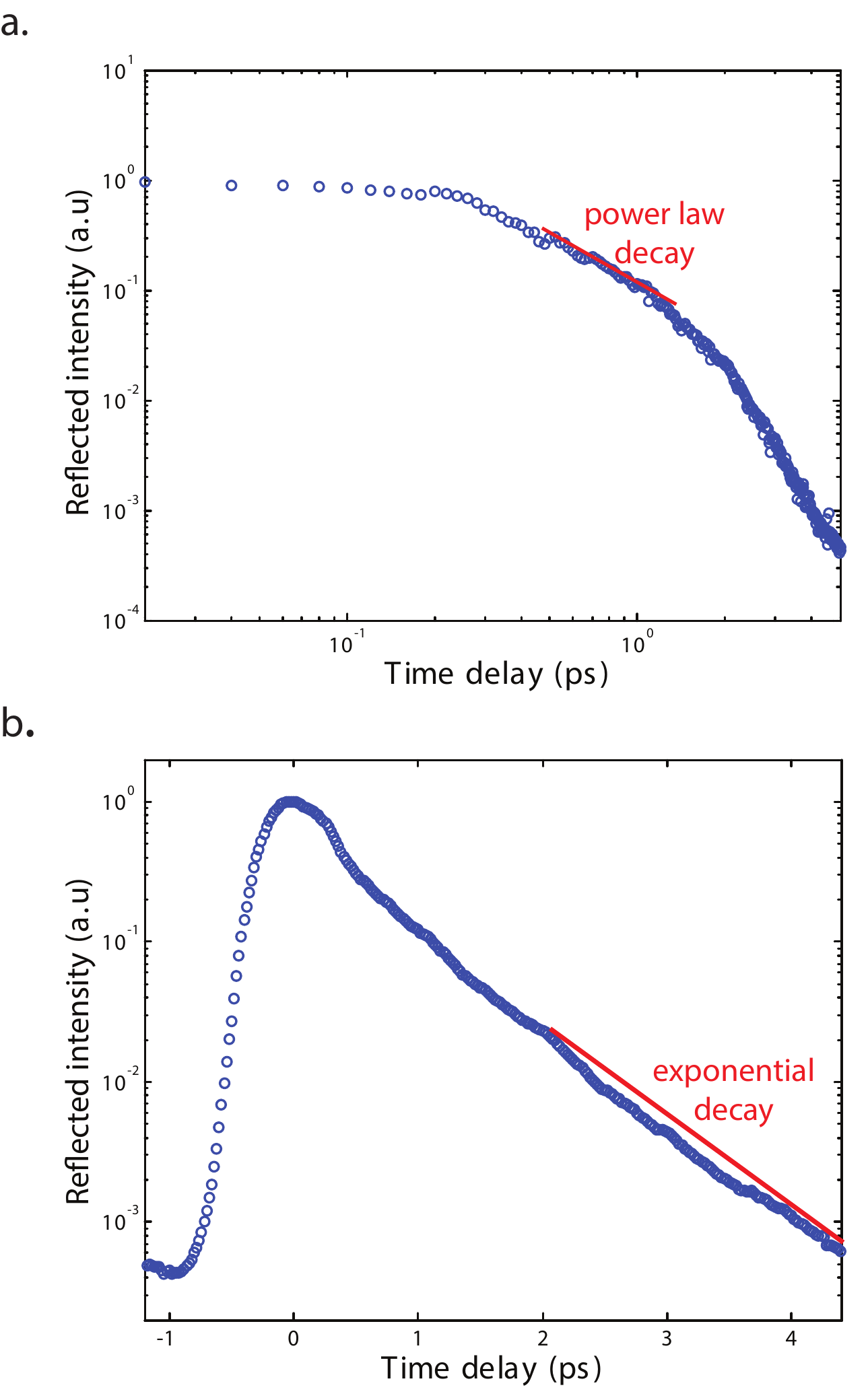}    
\caption{\textbf{Time-of-flight distribution of the intensity reflected by the TiO$_2$ sample}. (a) The time-evolution of $R(t)$ (blue circles) is plotted in a log-log scale and fitted with the $t^{-3/2}$ power law over the time range [500 fs - 1 ps]. (b) The time-evolution of $R(t)$ (blue circles) is plotted in a linear-log scale and fitted with an exponential decay over the time range [2 ps - 4.5 ps].}
\label{time}
\end{center}
\end{figure}

An experimental set up involving a coherent illumination scheme and an interferometric arm in reception \cite{badon2} has been used to measure $R(t)$ in the TiO$_2$ layer over an extensive range of time-of-flight. The result is displayed in Fig.\ref{time}. As predicted by theory [Eq.\ref{power}], the experimental time-of-flight distribution can be qualitatively fitted by the $t^{-3/2}$ power law for $t<<\tau_D$ [see  Fig.\ref{time}(a)]. 
In agreement with theory, the time decay of $R(t)$ becomes exponential at larger times of flight [see  Fig.\ref{time}(b)]. If we neglected absorption losses, a fit of the exponential decay would yield an estimation of the diffusion constant $D\sim$1250 m$^2$.s$^{-1}$. This value is quite far from our measurement deduced with the growth of the diffusive halo ($D\simeq$ 730 $m^2$.s$^{-1}$, see Sec.3A of the accompanying paper). The mismatch between both values can be easily accounted for by the presence of absorption losses. Considering that $D\simeq$ 730 $m^2$.s$^{-1}$, the exponential decay fit performed in Fig.\ref{time}(b) leads to the following estimation for the absorption mean free path: $l_a \simeq 118$ $\mu$m. This time-of-flight experiment illustrates the benefit of our approach that allows a quantitative measurement of the diffusion constant, independent from the absorption losses.


\end{document}